\documentclass[a4paper,11pt]{article}
\usepackage{pos}
\usepackage{lineno}

\usepackage{xspace}
\newcommand{\BdJpsiKS}{\ensuremath{B_d^0\to J/\psi K^0_{\text{S}}}\xspace}
\newcommand{\BsJpsiKS}{\ensuremath{B_s^0\to J/\psi K^0_{\text{S}}}\xspace}
\newcommand{\BdJpsiPi}{\ensuremath{B_d^0\to J/\psi\pi^0}\xspace}
\newcommand{\BsJpsiPhi}{\ensuremath{B_s^0\to J/\psi\phi}\xspace}
\newcommand{\BdJpsiRho}{\ensuremath{B_d^0\to J/\psi\rho^0}\xspace}
\newcommand{\BJpsiX}{\ensuremath{B_q^0\to J/\psi X}\xspace}

\title{Penguin Effects in \BdJpsiKS and \BsJpsiPhi}

\author[a]{Marten Z. Barel}
\author*[a,b]{Kristof De Bruyn}
\author[a,c]{Robert Fleischer}
\author[a]{Eleftheria Malami}

\affiliation[a]{Nikhef,\\ Science Park 105, 1098 XG Amsterdam, Netherlands}
\affiliation[b]{Van Swinderen Institute for Particle Physics and Gravity, University of Groningen,\\
9747 Groningen, Netherlands}
\affiliation[c]{Faculty of Science, Vrije Universiteit Amsterdam,\\
1081 HV Amsterdam, Netherlands}

\emailAdd{k.a.m.de.bruyn@rug.nl}

\abstract{
Controlling the contributions from doubly Cabibbo-suppressed penguin topologies in the decays \BdJpsiKS and \BsJpsiPhi is mandatory to reach the highest possible precision in the measurement of the $B^0_q$--$\bar B^0_q$ ($q=d,s$) mixing phases $\phi_d$ and $\phi_s$.
The penguin contributions can be determined using a strategy based on the $SU(3)$ flavour symmetry of QCD.
Using the latest experimental data, we update our combined analysis of the decays \BdJpsiKS, \BsJpsiPhi and their control channels \BsJpsiKS, \BdJpsiPi and \BdJpsiRho.
This allows us to simultaneously determine the penguin parameters and both mixing phases.
We discuss how the branching fractions of these decays can be used to probe the size of non-factorisable $SU(3)$-breaking effects, which form the main theoretical uncertainty associated with our $SU(3)$-based strategy, and provide new insights into the factorisation approach.
}

\FullConference{%
  11th International Workshop on the CKM Unitarity Triangle (CKM2021)\\
  22-26 November 2021\\
  The University of Melbourne, Australia
}

\begin{document}
\maketitle

\section{Introduction}

High precision measurements of the CP-violating phases $\phi_d$ and $\phi_s$, associated with mixing between neutral $B_q^0$ and $\bar B_q^0$ mesons $(q=d,s)$, offer excellent opportunities to search for signs of physics beyond the Standard Model (SM).
The discovery of New Physics (NP) contributions to the phases $\phi_q$ relies both on improved experimental measurements, and equally small theoretical uncertainties associated with the interpretation of these results.
The former can be expected from Belle II and the experiments participating in the high-luminosity phase of the Large Hadron Collider (HL-LHC). 
To achieve the latter, it is necessary to control contributions from higher-order decay topologies, which are often still neglected today, in all the decay channels used to measure $\phi_d$ and $\phi_s$.
In particular, this applies to the doubly Cabibbo-suppressed penguin topologies affecting the decay channels \BdJpsiKS and \BsJpsiPhi, which are considered the golden modes for the determination of $\phi_d$ and $\phi_s$, respectively.
Due to the presence of these penguin topologies, the CP asymmetries in \BdJpsiKS and \BsJpsiPhi only allow us to determine effective mixing phases $\phi_q^{\text{eff}}$, which are related to $\phi_q$ via hadronic shifts $\Delta\phi_q$.
The sizes of the penguin shifts $\Delta\phi_q$ are of the same order as the current experimental uncertainties to $\phi_q^{\text{eff}}$, and thus will become the dominant sources of systematic uncertainty in the determination of $\phi_d$ and $\phi_s$ if penguin effects remain unaccounted for.
These proceedings summarise the results of a strategy to determine the penguin shifts $\Delta\phi_d$ in \BdJpsiKS and $\Delta\phi_s$ in \BsJpsiPhi, employing the $SU(3)$ flavour symmetry of QCD.
They are based on the analysis presented in Ref.\ \cite{Barel:2020jvf}, but numerical results have been updated for the CKM 2021 conference.

\section{Determination of the Penguin Parameters}

The direct and mixing-induced CP asymmetries, $\mathcal{A}_{\text{CP}}^{\text{dir}}$ and $\mathcal{A}_{\text{CP}}^{\text{mix}}$, in \BdJpsiKS and \mbox{\BsJpsiPhi} allow us to determine the effective mixing phases $\phi_q^{\text{eff}}$:
\begin{equation}
     \frac{\eta_f \mathcal{A}_{\text{CP}}^{\text{mix}}(B_q\to f)}{\sqrt{1 - \left(\mathcal{A}_{\text{CP}}^{\text{dir}}(B_q\to f)\right)^2}}
     = \sin\left(\phi_q^{\text{eff}}\right)
     = \sin\left(\phi_q^{\text{SM}} + \phi_q^{\text{NP}} + \Delta\phi_q\right)\:,
\end{equation}
where $\eta_f$ is the CP eigenvalue of the final state $f$.
These quantities differ from the $B_q^0$--$\bar B_q^0$ mixing phases $\phi_q = \phi_q^{\text{SM}} + \phi_q^{\text{NP}}$ through the shifts $\Delta\phi_q$ associated with the penguin topologies.
Controlling the $\Delta\phi_q$ to a precision similar to that of the experimental measurements of $\phi_q^{\text{eff}}$ is mandatory to find new physics in neutral $B_q$ meson mixing in the future.

Here we summarise a strategy utilising the $SU(3)$ flavour symmetry of QCD to determine the penguin shifts $\Delta\phi_d$ and $\Delta\phi_s$ in \BdJpsiKS and \BsJpsiPhi, respectively.
Further details regarding this analysis can be found in Ref.\ \cite{Barel:2020jvf}.
The \BdJpsiKS decay amplitude can be written in the form
\begin{equation}\label{eq:Amp_BdJpsiKS}
    A(B_d^0\to J/\psi K^0) = \left(1-\frac{1}{2}\lambda^2\right)\mathcal{A}'\left[1+\epsilon a'e^{i\theta'}e^{i\gamma}\right]\:,\qquad
    \epsilon \equiv \frac{\lambda^2}{1-\lambda^2} \approx 0.052\:,
\end{equation}
where $\lambda \equiv |V_{us}|$ is an element of the Cabibbo--Kobayashi--Maskawa (CKM) quark mixing matrix; $\mathcal{A}'$ represents the contribution from the tree topology; $a'$ is the contribution from the penguin topologies relative to the tree topology; $\theta'$ is the associated strong phase difference; and $\gamma$ one of the angles of the Unitarity Triangle (UT).
The penguin parameters $a'$ and $\theta'$ are defined as
\begin{equation}
    ae^{i\theta} \equiv R_b\left[\frac{\text{Pen}^{(u)}-\text{Pen}^{(t)}}{\text{Tree}+\text{Pen}^{(c)}-\text{Pen}^{(t)}}\right]\:,
\end{equation}
where Pen$^{(q)}$ represents the penguin topology with internal quark flavour $q$, and $R_b$ is one of the UT sides.
The factor $\epsilon$ in the amplitude \eqref{eq:Amp_BdJpsiKS} suppresses the contributions from the penguin topologies, but their presence nonetheless affect the determination of $\phi_d$.
The CP asymmetries and the penguin shift $\Delta\phi_d$ can be expressed in terms of the penguin parameters $a'$ and $\theta'$ and the UT angle $\gamma$, where the expressions are given in Ref.\ \cite{Barel:2020jvf}.
The decay amplitude of the \BsJpsiPhi decay can we written in a similar form as Eq.\ \eqref{eq:Amp_BdJpsiKS} when neglecting the exchange and penguin-annihilation topologies, which are expected to be even smaller than the penguin contributions.

The penguin parameters can be determined from decay channels that are related to \mbox{\BdJpsiKS} via the $SU(3)$ flavour symmetry.
For example, the decay \BsJpsiKS is related to \BdJpsiKS through the interchange of all down and strange quarks, resulting in a one-to-one correspondence between all decay topologies of both channels.
The \BsJpsiKS decay amplitude can be written in the form
\begin{equation}\label{eq:Amp_BsJpsiKS}
    A(B_s^0\to J/\psi K_{\text{S}}^0) = -\lambda\mathcal{A}\left[1-ae^{i\theta}e^{i\gamma}\right]\:.
\end{equation}
Here, the contribution of the penguin topologies is not suppressed relative to the tree topology.
However, the overall decay amplitude is suppressed compared to \BdJpsiKS, thereby making this decay experimentally more challenging to study.
A similar expression as Eq.\ \eqref{eq:Amp_BsJpsiKS} can be given for the \BdJpsiPi and \BdJpsiRho decays, omitting exchange and penguin-annihilation topologies.

The strategy to determine $\Delta\phi_q$ is as follows:
The measured CP asymmetries in the penguin control modes \BsJpsiKS, \BdJpsiPi and \BdJpsiRho are combined with external input on the UT angle $\gamma = (64.9 \pm 4.5)^{\circ}$ \cite{LHCb:2021dcr} to determine the penguin parameters $a$ and $\theta$ in a theoretically clean way.
Next, the $SU(3)$ flavour symmetry, which implies
\begin{equation}\label{eq:su3_relation}
a' = a\qquad \text{and} \qquad \theta'=\theta\:,
\end{equation}
is used to relate the penguin parameters $(a,\theta)$ in the control modes to their counterparts $(a',\theta')$ affecting \BdJpsiKS and \BsJpsiPhi.
Knowing $a'$ and $\theta'$, the penguin shift $\Delta\phi_q$ can be calculated, and the $B_q^0$--$\bar B_q^0$ mixing phase $\phi_q$ obtained from the effective mixing phase $\phi_q^{\text{eff}}$.

\begin{figure}
    \centering
    \includegraphics[width=0.6\textwidth]{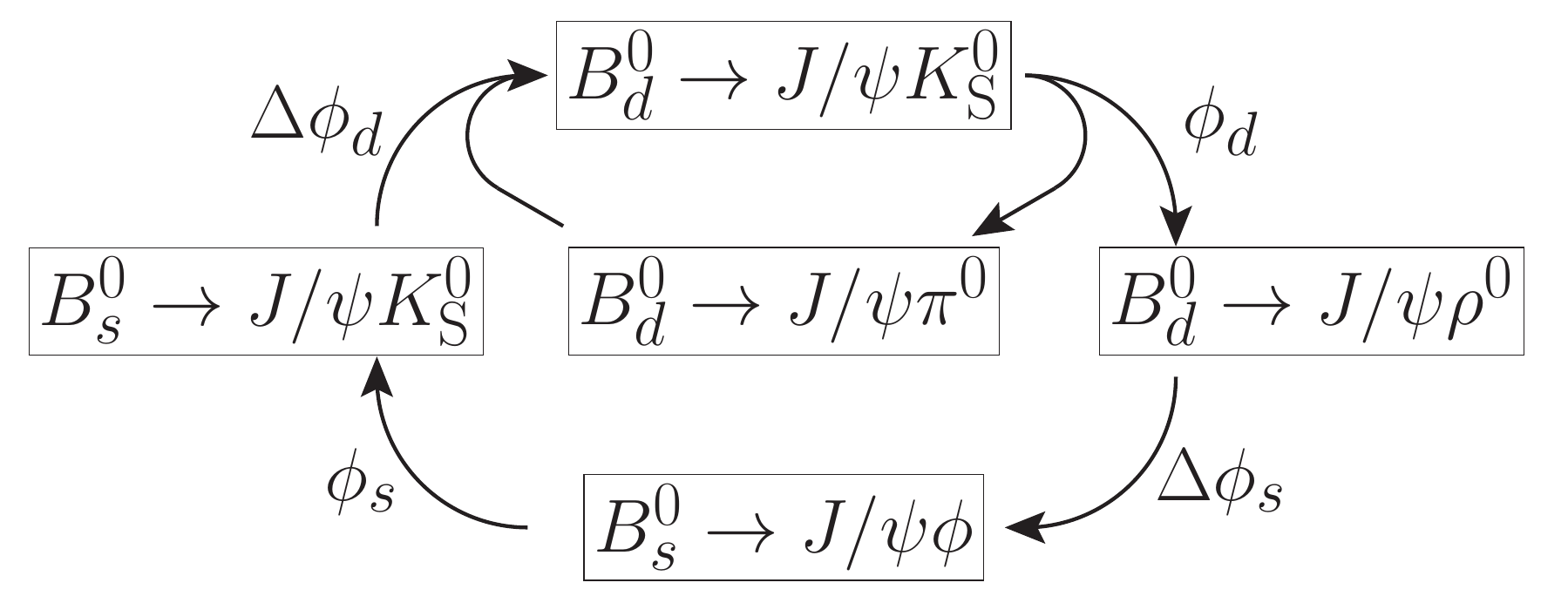}
    \caption{The cross-dependence between the determination of $\phi_d$ and $\phi_s$ and their penguin shifts, showing the interplay between the five \BJpsiX decays discussed here (taken from Ref.\ \cite{Barel:2020jvf})}.
    \label{fig:Interplay}
\end{figure}

There is, however, one additional complication, as illustrated in Fig.\ \ref{fig:Interplay}.
The mixing-induced CP asymmetry in \BsJpsiKS, needed to determine $a'$ and $\theta'$ in \BdJpsiKS, depends on the  $B_s^0$--$\bar B_s^0$ mixing phase $\phi_s$.
Likewise, additional information on the $B_d^0$--$\bar B_d^0$ mixing phase $\phi_d$ is required to determine the penguin parameters in \BdJpsiPi and \BdJpsiRho.
To take the dependencies between $\phi_d$, $\Delta\phi_d$, $\phi_s$ and $\Delta\phi_s$ into account, we have presented a strategy in Ref.\ \cite{Barel:2020jvf} involving all five decay channels.

\begin{figure}
    \centering
    \includegraphics[width=0.45\textwidth]{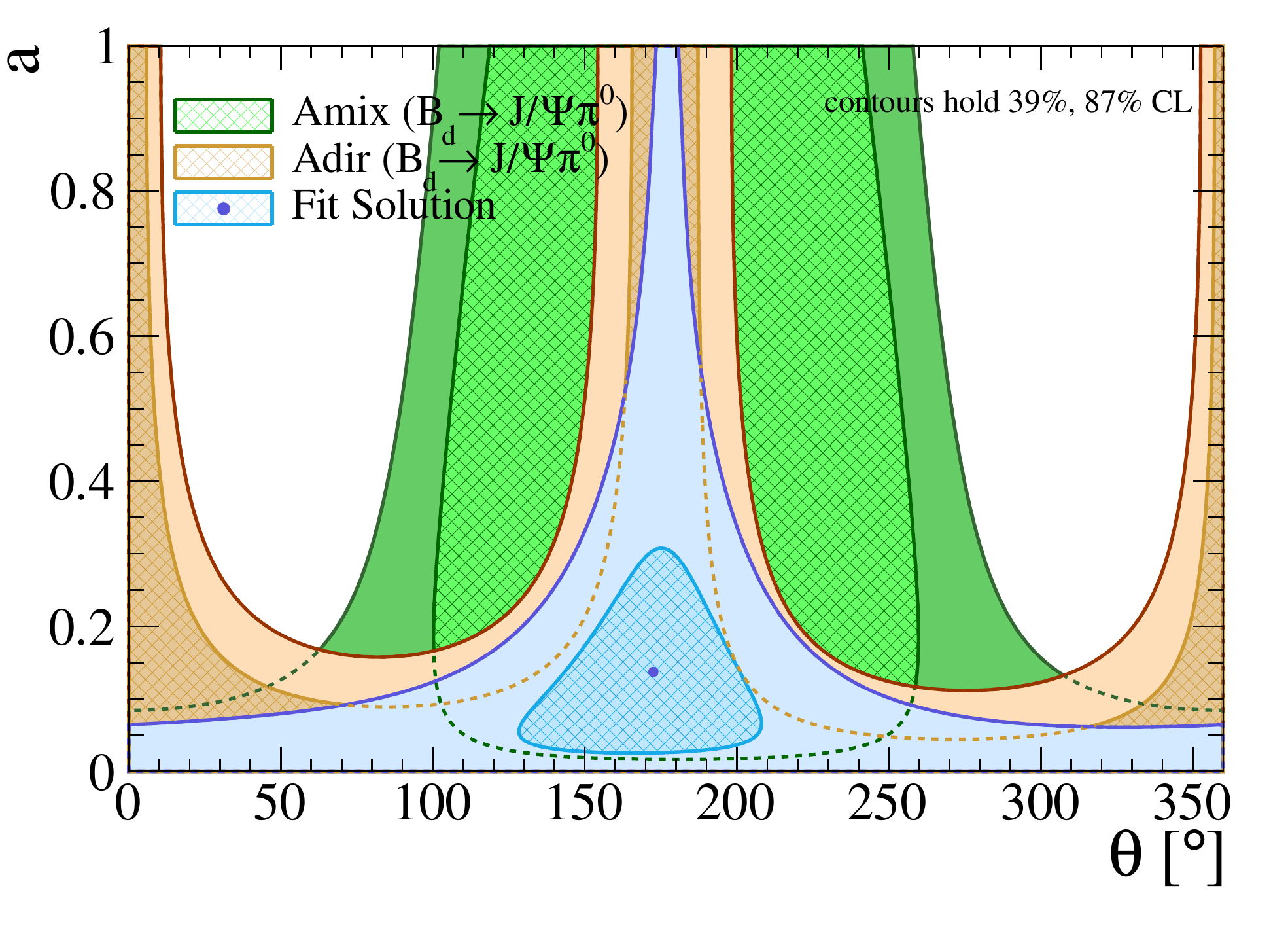}
    \includegraphics[width=0.45\textwidth]{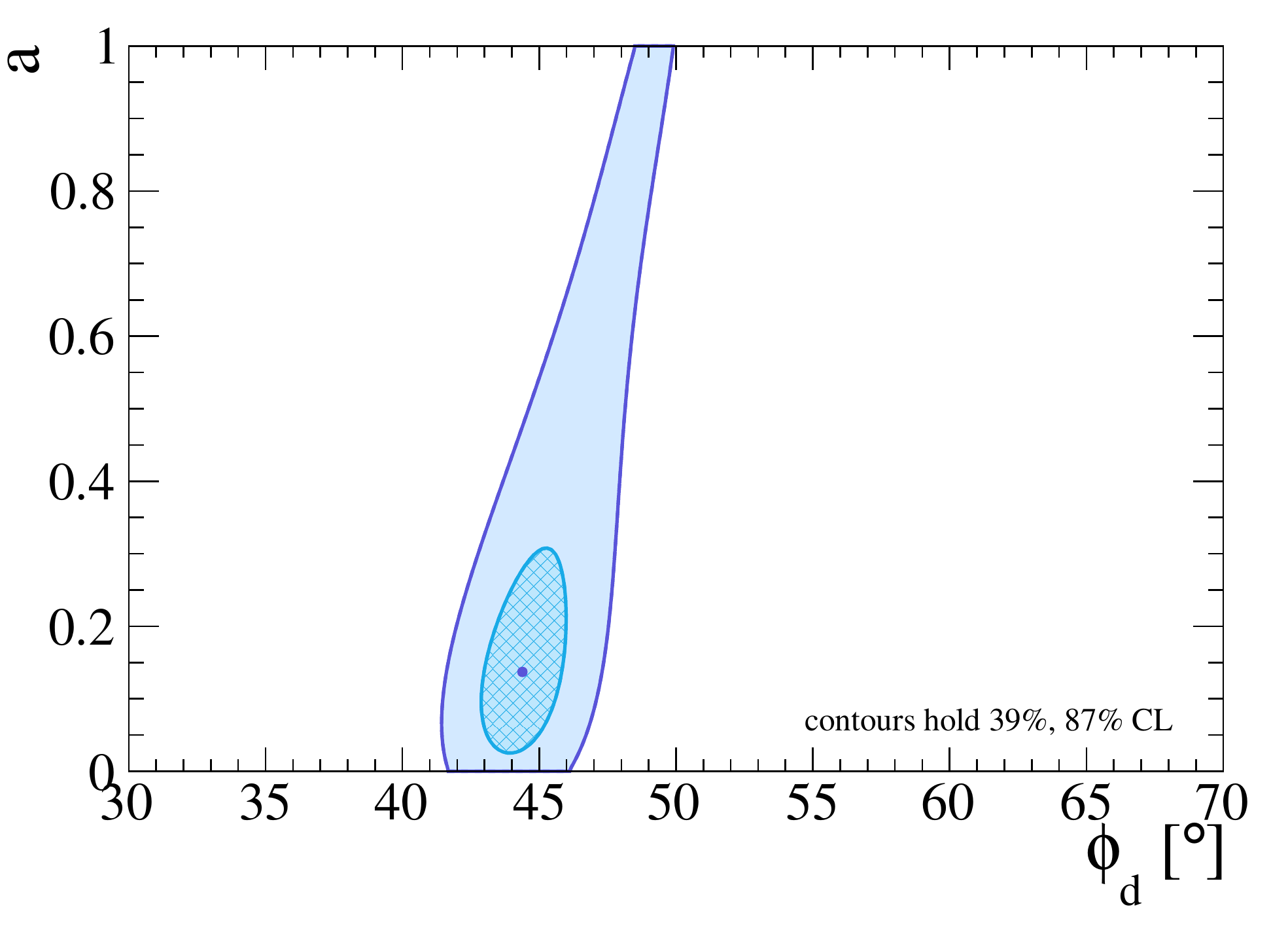}
    
    \includegraphics[width=0.45\textwidth]{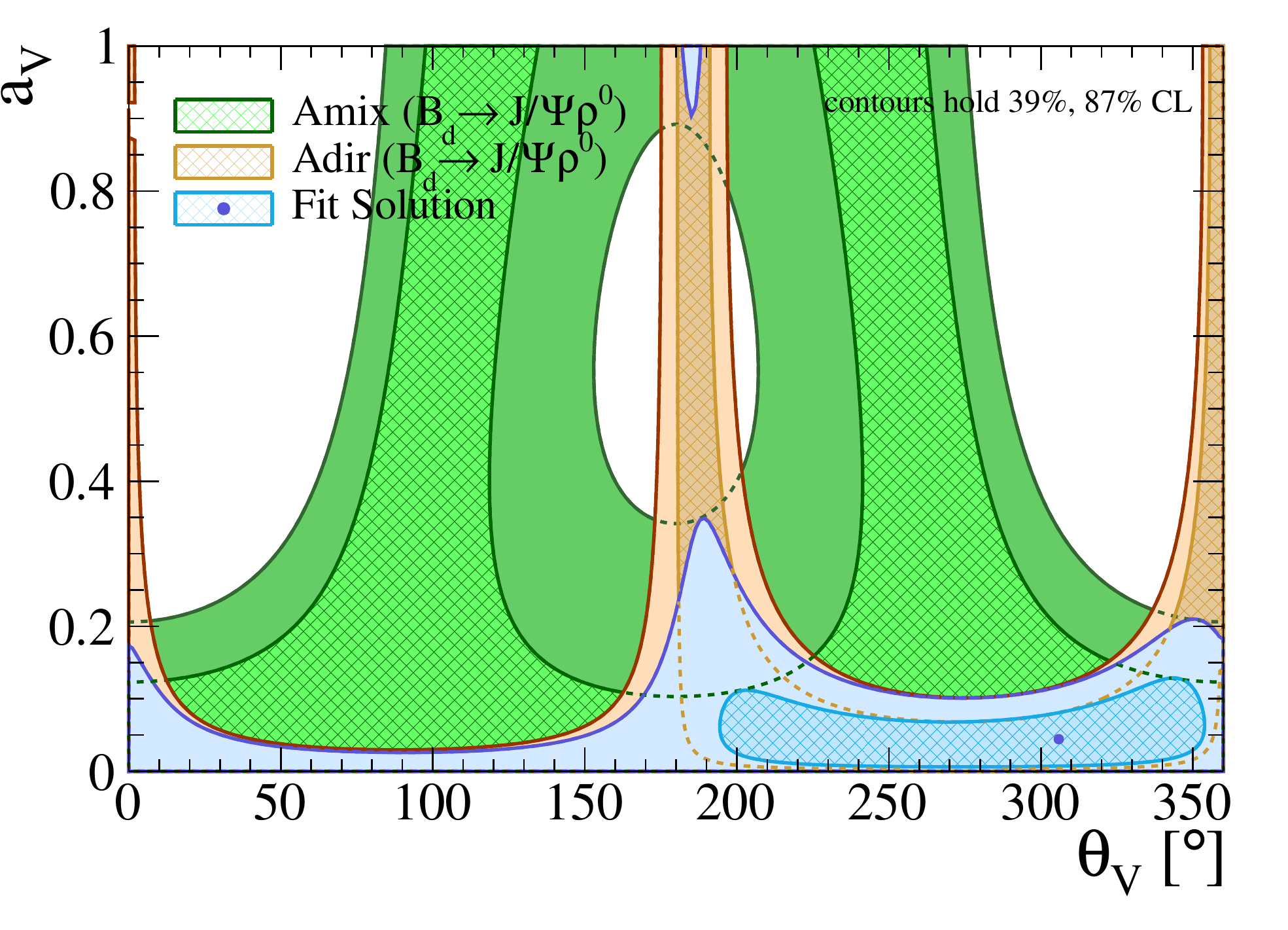}
    \includegraphics[width=0.45\textwidth]{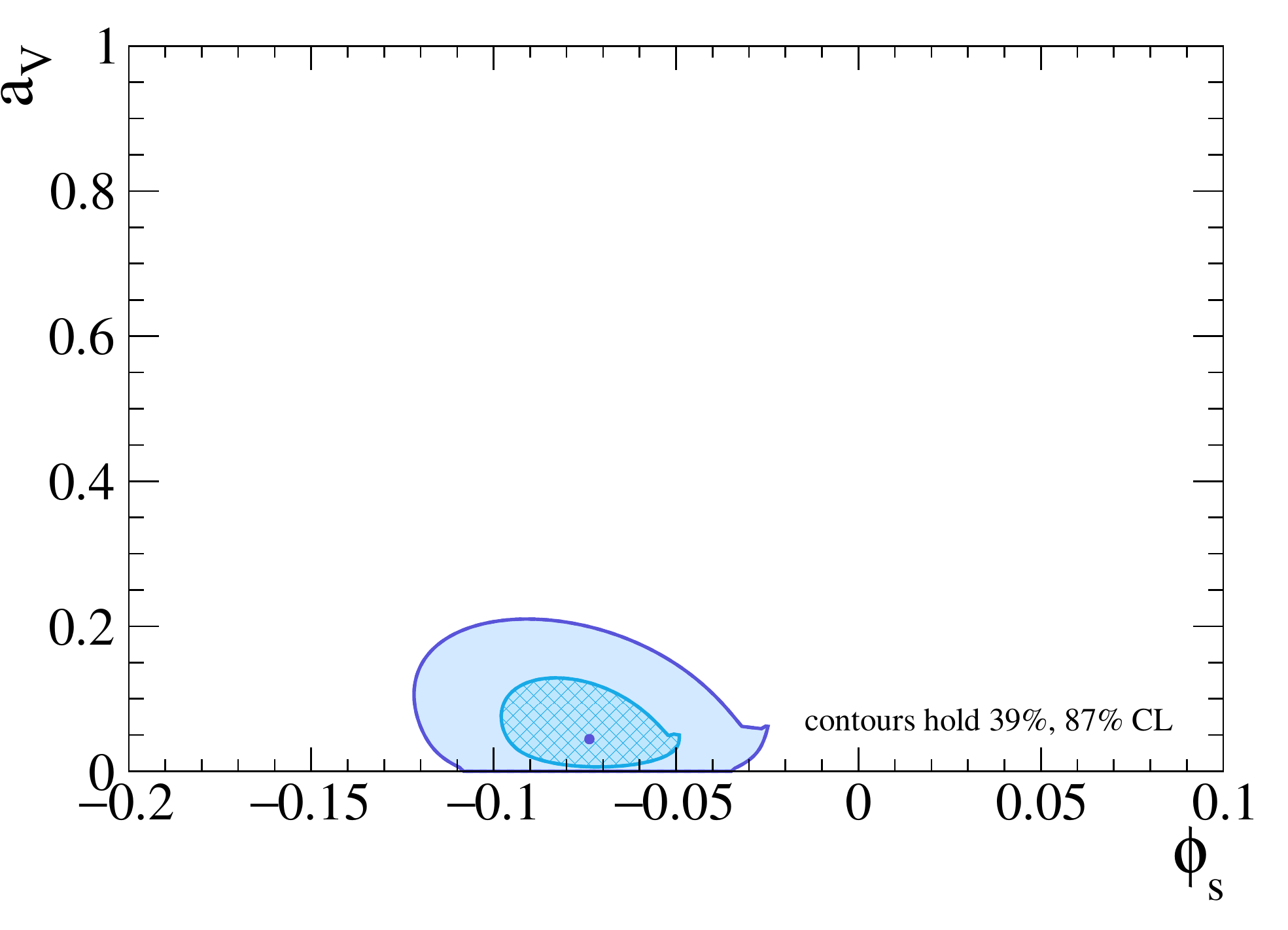}
    \caption{Two-dimensional confidence regions of the fit for the penguin parameters, $\phi_d$ and $\phi_s$ from the CP asymmetries in the \BJpsiX decays.
    The contours for $\mathcal{A}_{\text{CP}}^{\text{dir}}$ and $\mathcal{A}_{\text{CP}}^{\text{mix}}$ are added for illustration only.
    They include the best fit solutions for $\phi_d$, $\phi_s$ and $\gamma$ as Gaussian constraints.
    }
    \label{fig:MasterFit}
\end{figure}

The updated results of this simultaneous analysis for the penguin parameters affecting \BdJpsiKS, \mbox{\BsJpsiKS} and \BdJpsiPi are given as follows:
\begin{equation}\label{eq:results_phid}
    a = 0.14_{-0.11}^{+0.17}\:, \qquad
    \theta = \left(173_{-45}^{+35}\right)^{\circ}\:, \qquad
    \phi_d = \left(44.4_{-1.5}^{+1.6}\right)^{\circ}\:.
\end{equation}
Comparing the fit value of $\phi_d$ with the experimental input $\phi_{d,J/\psi K^0}^{\text{eff}} = \left(43.6 \pm 1.4\right)^{\circ}$ shows the non-negligible impact of the penguin topologies.
The two-dimensional confidence regions for $a$, $\theta$ and $\phi_d$ are shown in Fig.\ \ref{fig:MasterFit}.
The updated results for the penguin parameters affecting \BsJpsiPhi and \BdJpsiRho take the following values:
\begin{equation}\label{eq:results_phis}
    a_V = 0.044_{-0.038}^{+0.085}\:, \qquad
    \theta_V = \left(306_{-112}^{+\phantom{0}48}\right)^{\circ}\:, \qquad
    \phi_s = -0.074_{-0.024}^{+0.025} = \left(-4.2 \pm 1.4\right)^{\circ}\:.
\end{equation}
Also here, the impact of the penguin topologies can be seen by comparing the fit value of $\phi_s$ with the experimental input $\phi_{s,J/\psi\phi}^{\text{eff}} = -0.071 \pm 0.022 = (-4.1 \pm 1.3)^{\circ}$.
The two-dimensional confidence regions for $a_V$, $\theta_V$ and $\phi_s$ are shown in Fig.\ \ref{fig:MasterFit}.

\begin{figure}
    \centering
    \includegraphics[width=0.45\textwidth]{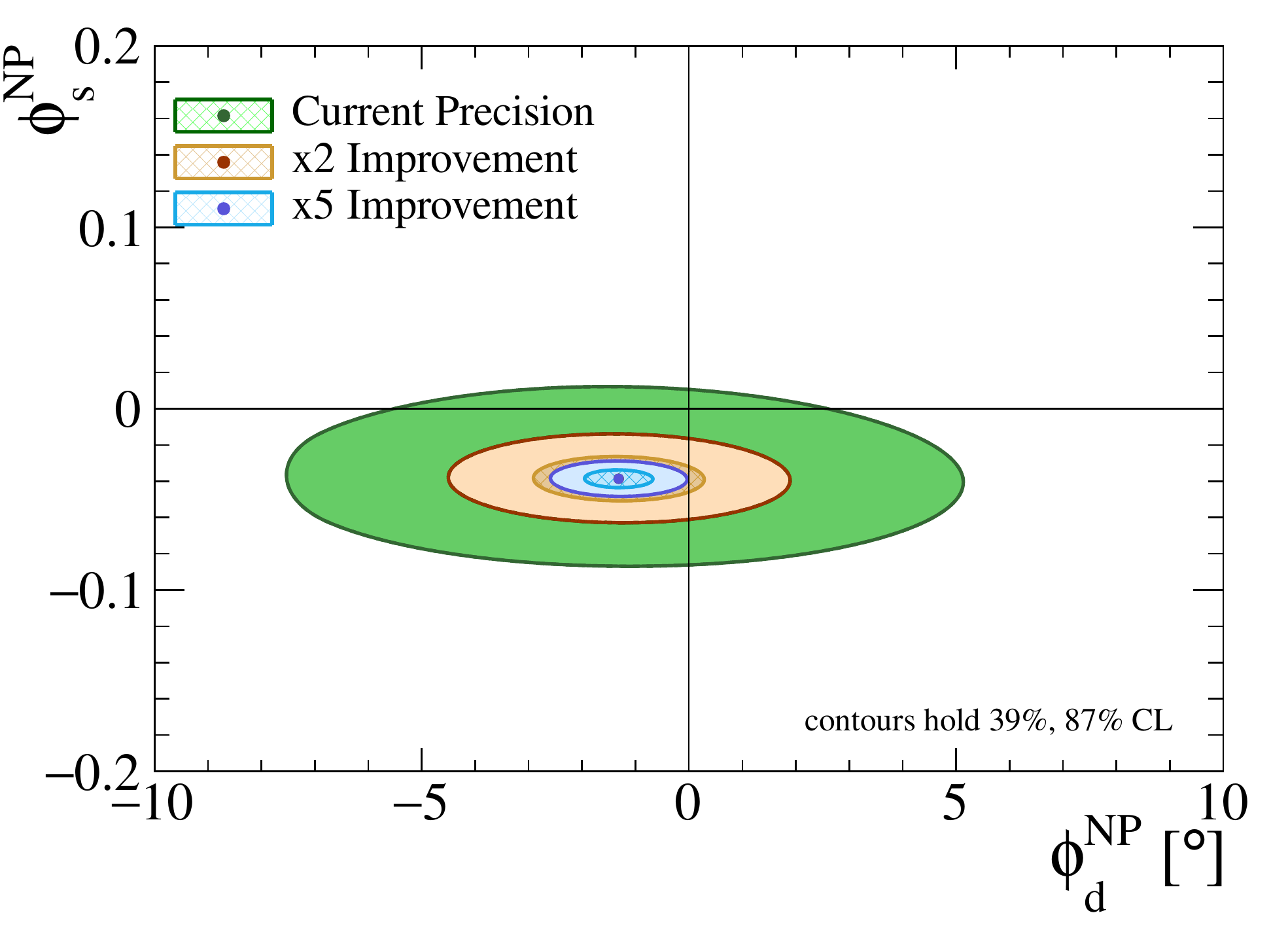}
    \caption{Comparison between the current precision of the NP phases $\phi_d^{\text{NP}}$ and $\phi_s^{\text{NP}}$ with two future benchmark scenarios in which we assume an overall improvement of the experimental input measurements by a factor 2 and 5.
    Shown are the two-dimensional confidence regions of the fit.}
    \label{fig:future_NP}
\end{figure}

Combining the result in Eq.\ \eqref{eq:results_phis} with the SM prediction
\begin{equation}
    \phi_s^{\text{SM}} = -0.0351 \pm 0.0021 = (-2.01 \pm 0.12)^{\circ}\:,
\end{equation}
which is based on a fit of the unitarity triangle using only $\gamma$ and $R_b$, we find a NP phase
\begin{equation}
    \phi_s^{\text{NP}} = -0.039 \pm 0.025 = (-2.2 \pm 1.4)^{\circ}\:.
\end{equation}
Although this result is compatible with zero below the $2\,\sigma$ level, it leads to interesting prospects for the Belle II and HL-LHC era.
The benchmark scenarios in Fig.\ \ref{fig:future_NP} illustrate that discovering NP in $B_s$ mixing is still possible, but only if improvements are made to the CP asymmetry measurements in all five \BJpsiX decays.
On the other hand, finding NP in $B_d$ mixing is limited by the knowledge of the UT apex.

\section{Determination of the Hadronic Parameters}

\begin{figure}
    \centering
    \includegraphics[width=0.45\textwidth]{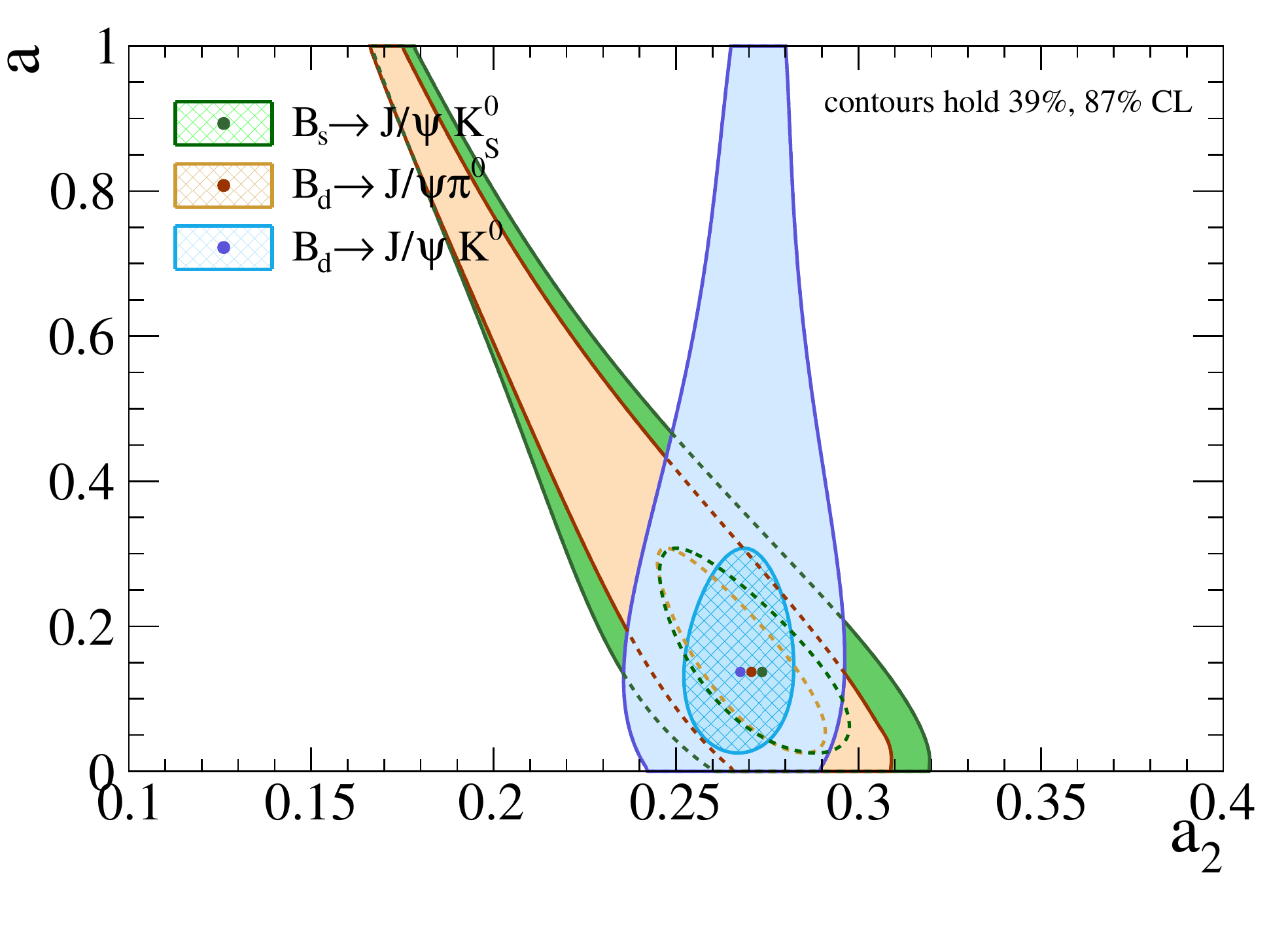}
    \includegraphics[width=0.45\textwidth]{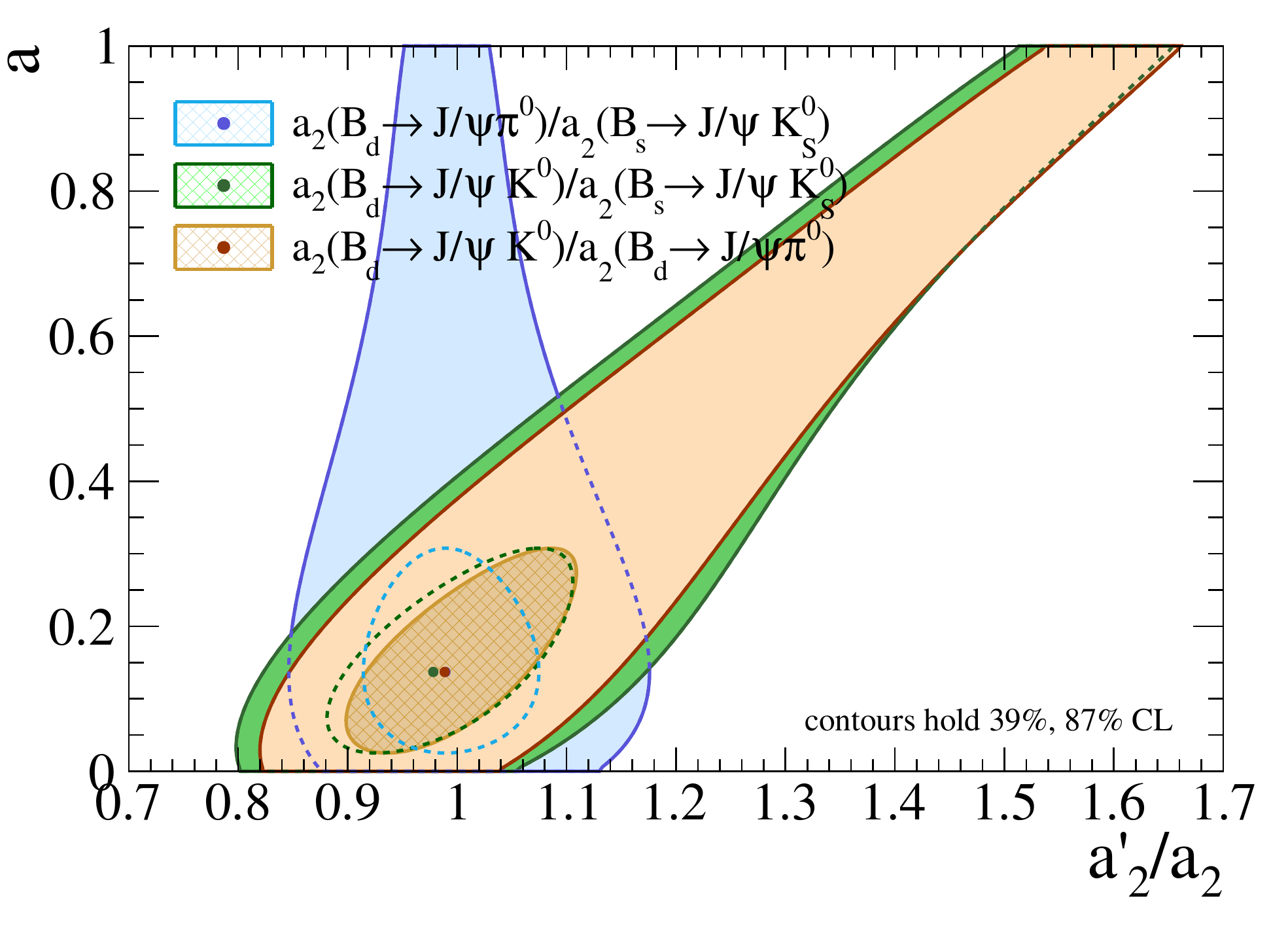}
    \caption{Two-dimensional confidence regions for the effective colour-suppression factors and their ratios.}
    \label{fig:a2_factor}
\end{figure}

The main theoretical uncertainties in the determination of the penguin shifts $\Delta\phi_q$ arise from the breaking of the $SU(3)$ flavour symmetry, which implies that relation \eqref{eq:su3_relation} is not exact.
Contributions to $SU(3)$-breaking come both from factorisable and from non-factorisable effects.
The ratio between the kaon and pion decay constants, which differs from one by about 20\% \cite{Aoki:2021kgd}, gives a good estimate for the size of the factorisable effects.
However, because we are specifically looking at a ratio of amplitudes in Eq.\ \eqref{eq:Amp_BdJpsiKS}, the factorisable effects, which equally affect the tree and penguin topologies, drop out.
The penguin parameters $a$ and $\theta$ are therefore only affected by non-factorisable $SU(3)$ breaking, which is expected to be suppressed well below the 20\% level.
The size of these effects can be explored using the measured branching fractions.
For example, the branching fraction of the decay \BdJpsiPi can be written as

\begin{align}
   2 \: \mathcal{B}(B_d^0\to J/\psi\pi^0) =  & 
   \: \tau_{B_d} \: \frac{G_{\mathrm F}^2}{32 \pi} |V_{cd}V_{cb}|^2 \: m_{B_d}^3
   \left[ f_{J/\psi}  f_{B_d\to \pi}^+(m_{J/\psi}^2) \right]^2
    \left[\Phi\left(\frac{m_{J/\psi}}{m_{B_d}},\frac{m_{\pi^0}}{m_{B_d}}\right)\right]^3 \nonumber \\
    & \times (1 - 2 a\cos\theta\cos\gamma + a^2) \times \left[ a_2 (B_d^0\to J/\psi\pi^0) \right]^2,
\end{align}
where $G_{\mathrm F}$ is the Fermi constant; $f_{J/\psi}$ is the $J/\psi$ decay constant; $f_{B_d\to \pi}^+$ the $B_d\to\pi$ form factor; $\Phi$ a phase-space factor; and $a_2$ an effective colour-suppression factor that includes the non-factorisable corrections.
Combining the branching fraction measurement with the fit results \eqref{eq:results_phid} for the penguin parameters, we can determine the factor $a_2$ for the decays \BsJpsiKS, \BdJpsiPi and \BdJpsiKS, as illustrated in Fig.\ \ref{fig:a2_factor}, where the uncertainty is dominated by the lattice calculation of the form factor \cite{Aoki:2021kgd}.
The dependence on the form factor can be avoided by constructing a ratio between the \BJpsiX branching fraction and a suited semileptonic $B$ decay \cite{Barel:2020jvf}, greatly improving the constraint on $a_2$.

The obtained results in Fig.\ \ref{fig:a2_factor} are in good agreement with the naive expectation from factorisation, $a_2 = 0.21 \pm 0.05$ \cite{Buras:1998us}, showing deviations at the 30\% to 40\% level.
Although this estimate may appear to be large, it is a non-trivial result, as factorisation is not expected to work well in this family of decays.
Putting both effects together, non-factorisable $SU(3)$-breaking effects can thus be expected at a level of 5\% to 8\%.
This illustrates that the $SU(3)$-flavour strategy based is robust.
The impact of non-factorisable $SU(3)$-breaking effects is also illustrated by the ratios of $a_2$ parameters, shown in the right-hand panel of Fig.\ \ref{fig:a2_factor}.
In the absence of any non-factorisable $SU(3)$-breaking effects, these ratios are expected to be one.

\section{Conclusion}

From a combined analysis of the decays \BdJpsiKS, \BsJpsiPhi and their control channels \BsJpsiKS, \BdJpsiPi and \BdJpsiRho, the $B_q^0$--$\bar B_q^0$ mixing phases $\phi_d$ and $\phi_s$ were determined, taking into account the impact of penguin topologies on the measured CP asymmetries.


\end{document}